\documentclass[letterpaper, 10 pt, conference]{ieeeconf}  
\IEEEoverridecommandlockouts                              % This command is only needed if 
\usepackage{graphics} % for pdf, bitmapped graphics files
\usepackage{epsfig} % for postscript graphics files
\usepackage{mathptmx} % assumes new font selection scheme installed
\usepackage{times} % assumes new font selection scheme installed
\usepackage{amsmath} % assumes amsmath package installed
\usepackage{amssymb}  % assumes amsmath package installed
\usepackage{xcolor}
\usepackage[noadjust]{cite}
\usepackage{framed}
\usepackage{mdframed}

\title{\LARGE \bf Using Knowledge Awareness to improve Safety of Autonomous Driving}
\author{Andrea Calvagna$^{1}$, Arabinda Ghosh$^{2}$ and Sadegh Soudjani$^{2,3}$% <-this % stops a space
\thanks{*This work is supported by the Horizon Europe EIC project SymAware under the grant agreement 101070802.}% <-this % stops a space
\thanks{$^{1}$ Dipartimento di Matematica e Informatica, University of Catania, Italy
        {\tt\small andreamario.calvagna@unict.it}}%
\thanks{$^{2}$ Max Planck Institute for Software Systems, Kaiserslautern, Germany
        {\tt\small arabinda@mpi-sws.org sadegh@mpi-sws.org}}%
\thanks{$^{3}$ School of Computing, Newcastle University, United Kingdom
        {\tt\small sadegh.soudjani@newcastle.ac.uk}}%
}

\begin{document}

\maketitle
\thispagestyle{empty}
\pagestyle{empty}
%%%%%%%%%%%%%%%%%%%%%%%%%%%%%%%%%%%%%%%%%%%%%%%%%%%%%%%%%%%%%%%%%%%%%%%%%%%%%%%%
\begin{abstract}
We present a method, which incorporates \emph{knowledge awareness} into the symbolic computation of discrete controllers for reactive cyber physical systems, to improve decision making about the unknown operating environment under uncertain/incomplete inputs.
Assuming an abstract model of the system and the environment, we translate the knowledge awareness of the operating context into linear temporal logic formulas and incorporate them into the system specifications to  synthesize a controller. The knowledge base is built upon an ontology model of the environment objects and behavioural rules, which includes also symbolic models of partial input features. The resulting symbolic controller support smoother, early reactions, which improves the security of the system over existing approaches based on incremental symbolic perception. A motion planning case study for an autonomous vehicle has been implemented to validate the approach, and presented results show significant improvements with respect to safety of state-of-the-art symbolic controllers for reactive systems.
%use a knowledge base at the plant to elaborate on partial/unknown environmental inputs and take decisions    
\end{abstract}
%%%%%%%%%%%%%%%%%%%%%%%%%%%%%%%%%%%%%%%%%%%%%%%%%%%%%%%%%%%%%%%%%%%%%%%%%%%%%%%
%\AG{General comment: These words are interchangeably used throughout the paper: unreliable/uncertain/incomplete/partial input. However, the respective meanings are not the same. Why not just use partial input, which is exactly what we considered? Unreliability/uncertainty are different research fields altogether}
\section{INTRODUCTION}
%done\Sadegh{Update the reference:``Reasoning about knowledge: a survey''.}
Safe autonomous driving in an urban environment is currently among the most sought-after and complex application of cyber physical systems (CPS), requiring continuous real-time exchange of information among the sensors, cameras, and controllers. Malfunctions at any stage will lead to serious repercussions. Thus, it is imperative to systematically nurture knowledge about the environment in the autonomous driving system which will further improve safety in a complex environment. 
To this aim, a common approach is to consider an abstract form of the sophisticated raw sensor data processing, applied for objects identification and tracking, and represent perception as symbolic inputs~\cite{kress2018synthesis,nilsson2015correct,schillinger2018simultaneous}. Symbolic computation techniques are then applicable to the design and synthesis of reactive controllers for these CPS~\cite{fainekos2005temporal,finucane2010ltlmop}. % based on temporal logic specifications.  
However, assuming a fully observable environment with flawless detection of its objects, often results in overly restrictive or unsafe behaviours, as the perception modules can only react based on the exact knowledge of the environmental inputs~\cite{fayyad2020deep,sahin2019multirobot,wang2020hyperproperties}. 
%In real-world scenarios, the perception gradually improve as the autonomous vehicle moves closer to its target, resulting in incremental layers of perception refinement~\cite{kamale2022cautious}. Nonetheless, adverse weather conditions, dynamic operating conditions, and limitations in the existing hardware may affect the quality of the perceived information, which are unavailable all at once. It is then of absolute value to improve the planning by processing also incomplete/partial inputs, in order to improve safety by anticipating safe behaviours and promoting the early reactions of the car. 
%\models   we focus on autonomous driving as example of cyber phisical system but the approach i sgeneral
% a sentens to connect symbolic omedl to reactive systems
%\vDash
The main research question is: \textbf{can we synthesize a symbolic controller for a reactive CPS that enhances safety of the motion planning in the presence of incomplete environment information?}

\begin{figure}[t]
    \centering
    \includegraphics[width=\columnwidth]{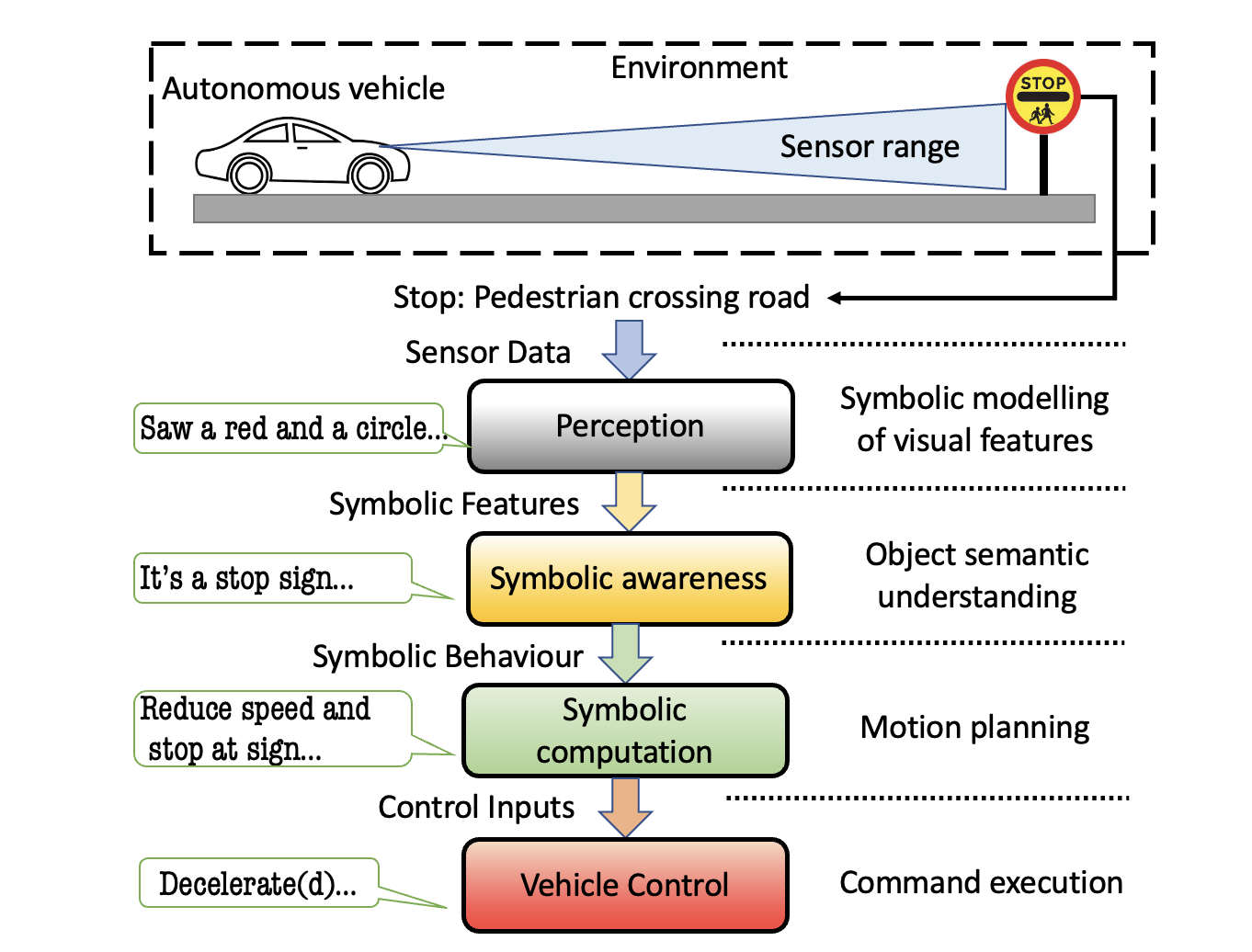}
    \caption{Modular architecture of the proposed reactive system. Motion planning layer leverages an enhanced model of symbolic inputs, including semantic awareness of environment objects and their required behavioral rules.} % motion planning as an example of CPS
    \label{fig:schema}
\end{figure}
%\subsection{Contribution} 
%anticipate figure before the controbution
\smallskip
\noindent\textbf{Contribution:} 
In this paper, an original approach to incorporate a \emph{knowledge base} into a reactive controller is proposed, encoding symbolic awareness of the \emph{entities} and \emph{rules} of its operating context. The layered architecture of the proposed solution is given in Fig.~\ref{fig:schema}. The \emph{knowledge base} is supplied as a set of linear temporal logic (LTL) specifications, supporting the controller’s decision-making process in the presence of partial inputs. Knowledge awareness of the operating context is obtained from a context \emph{ontology} model, formally describing the system interaction environment. This approach generalizes symbolic perception refinement, based on incremental perception models, and also is suitable to be streamlined as a completely automated reactive controller synthesis process. To the best of our knowledge, this is the first work incorporating a symbolic awareness knowledge base into the synthesis of a reactive controller. 

\smallskip 
The rest of this paper is organized as follows: after reviewing related work, the preliminaries and the problem statements are presented in section~\ref{sec:prel}. In section~\ref{sec:solution}, we presented the proposed approach to the automated synthesis of a knowledge-aware reactive controller. As a proof of concept, in section~\ref{sec:casestudy} we consider a small traffic-sign autonomous driving scenario as CPS and apply the proposed approach to synthesize a controller incorporating awareness of the \emph{semantics} of traffic signs. In section~\ref{sec:evaluation} we simulated the uncertainty of environmental inputs to evaluate the case study controller performance and show that it outperforms both basic and \emph{perception-tree} based controllers in early detecting and reacting to partial symbolic input features. Section~\ref{sec:conclusion} draws our conclusions and highlights future directions.

\noindent\textbf{Related Work:} Approaches to improve symbolic perception, supporting intermediate levels of detail, have been developed in the literature~\cite{fan2021multi, liang2020traffic}. Some research works have focused on developing the perception modules to estimate the attributes of the key environment components such as obstacles or other vehicles etc.~\cite{okumura2016challenges,gruyer2017perception}. %Other works either captured a notion of \emph{evolution} of the symbolic knowledge about the environment in terms of the \emph{scaling} of features that happens when getting closer to the object or just focused on safe motion planning with non-evolving, incomplete information~\cite{vasile2020reactive, ulusoy2014receding, fu2014abstractions}. % In this work, we instead focus on evolving perception scenarios, in terms of varying numbers of feature details. 
Kamale et al.~\cite{kamale2022cautious} have considered partial information produced by \emph{additive} symbolic perception modules and gradually refined the controller output for safer navigation. % A limitation of their approach appears to be its reliance on a strictly hierarchical model of perception refinement, % \emph{partitioning} all environment objects based on their own featured details (i.e. where no multiple inheritances were possible). Moreover, it 
%assuming that symbolic perception updates can happen in only one predefined ordered sequence. %As a consequence, different feature orderings or incomplete sets of additive features would be simply not understood by the controller. 
In contrast, we consider the problem of reactive planning without \emph{requiring} assumptions on the \emph{dynamics} of the symbolic inputs, so that incremental input refinement is subsumed as a special case of our approach. 

Another line of research aims at improving symbolic motion planning introducing ontology models, to describe and reason about environmental objects~\cite{staab2010handbook}, e.g. road signs, road networks and traffic-related concepts. An ontology-based context model was proposed by Fuchs et al.~\cite{fuchs2008integration} to represent knowledge for a human driver assistance system. Arai et al.~\cite{arai_accelerating_2021} applied machine learning algorithms to automatically extract a road sign ontology.
%, representing the salient visual features of a road signs that are discernible through sight or imaging.
Hulsen et al.~\cite{hulsen2011traffic} proposed an ontology-based framework to represent traffic rules and integrate reasoning among participating vehicles. %Zaji et al.~\cite{zaji2023ontology} introduced a knowledge base using ontologies and used an evolutionary algorithm to optimize traffic lights, to reduce total travel time. 
However, these works did not consider the knowledge base as also a model of partial information.  In contrast, our work is the first that considers an ontology model to derive a knowledge base, supporting partial environment information, and integrates it in the actual synthesis of a reactive controller. 
%%%%%%%%%%%%%%%%%%%%%%%%%%%%%%%%%%%%%%%%%%%%%%%%%%%%%%%%%%%%%%%%%%%%%%%%%%%%%%%%%%%%%%%

\section{Preliminaries and Problem Statement}
\label{sec:prel}
%\AG{Are we not explaining the notations? we had a notations section previously, I see now its deactivated!!} sadegh removed it
%\noindent\textbf{Notation.}
%For a set $A$, the power set of $A$ is denoted by $2^A$, which is the set of subsets of $A$. The finite set $\{1,2,\ldots,n\}$ is denoted by $[1..n]$. The cardinality of a finite set $A$ is denoted by $|A|$. %\Sadegh{check the notation throughout the paper, make sure consistent and well-defined.} 

\noindent\textbf{Symbolic Model:} Discrete controllers for symbolic motion planning in planar space environments require abstracting the continuous operating domain into symbolic models of space, inputs and outputs. In \emph{classic} designs, symbolic inputs/outputs are sets $U$ and $Y$ of atomic propositions abstracting the environmental \emph{objects} recognized by the perception modules sensors, and the \emph{actions} executable by the system continuous controller modules, respectively.
The continuous space domain is also abstracted as a set of discrete states of a finite transition system. The cell size of the discrete space model is assumed to guarantee the validity of the symbolic abstractions~\cite{alur2000discrete}. 

\noindent\textbf{Definition 1 (LTL specifications):} {\it Temporal logic specifications are logic formulas over a finite set of atomic propositions $\Pi$.
%$p_i \in \Pi$ with $i \in \{1,2,\ldots,|\Pi|\}$.
LTL syntax is recursively defined as: $\phi::=true|p|\lnot\phi|\phi_1\wedge\phi_2|\bigcirc\phi|\phi_1 {\mathcal{U}} \phi_2$, 
where true, false are Boolean constants; negation ($\lnot$), conjunction ($\wedge$) are Boolean operators;  next ($\bigcirc$) and until ($\mathcal{U}$) denote the temporal operators}. Additional logic and LTL operators such as disjunction ($\vee$), eventually ($\lozenge$), implication ($\Rightarrow$), always ($\square$), can be derived from these. We point the reader to~\cite{pnueliLTL77} for the semantic definition of the LTL operators. 
To constrain the system behaviours, we define temporal logic \emph{safety} and \emph{liveness} specifications as formulas composed of the symbolic outputs in $Y$, as atomic propositions, and the LTL operators. As an example, a safety specification could be, to never collide: $\square \lnot collide$; and a liveness specification, could be to reach a target infinitely often: $\square\lozenge (target \wedge \bigcirc move)$.

\noindent\textbf{Definition 1 (Finite Transition System):} {\it We model the system as a labelled finite transition system (FTS). An FTS is a tuple $T = (Q, \theta,\Sigma,\Pi,\tau, O)$, where $Q$ is a finite set of states; $\theta \subset Q$ is the set of possible start states; $\Sigma \subseteq Q\times Q$ is a set of state transitions; $\Pi := \{p_1 ,\dots ,p_n\}$ is a set of state properties modelled as atomic propositions;  $\tau:Q \rightarrow 2^{\Pi}$ is a labelling function, assigning properties to each state;  and $O$ is a function mapping each \emph{edge} $e::= q \rightarrow q\prime\;, s.t. (q,q\prime)\in\Sigma\;$, to a corresponding output $y \in Y$. 
}

An FTS describes a set of sequences of states, as all the paths through it. As the labels annotate the states along these paths, we constrain the set of legal (\emph{motion}) paths by specifying temporal logic formulas on those labels. A paths is also a sequence of edges $\bar e = e_0, e_1, \dots $ such that $e_k \in \Sigma\;, \forall~k \geq 0$, generating an output trajectory $\bar y = y_0, y_1 \dots \; $, where $y_k = O(e_k),\; \forall~k \geq 0$.  %We can express constraints on the set of safe \emph{behaviours} of the system by specifying temporal logic formulas on those outputs. Temporal logic specifications (LTL) are commonly used to express the safety constraints and the liveness goals for both the system and the environment. 

To model incomplete information about the environment we leverage the concept of \emph{awareness}, introduced in the literature about logic frameworks for symbolic knowledge representation and symbolic computation~\cite{Halpern95,Fagin1991AMA}.

\smallskip
\noindent\textbf{Definition 3 (Knowledge Awareness):} {\it We consider symbolic \emph{knowledge} of a \emph{fact} or property  as $M\vDash \phi\;$, where $\phi$ is the symbolic encoding of the property as a Boolean formula, evaluating true or false, and $M$ is a set of truth assignments to the atomic propositions in $\phi$. Knowledge is thus intuitively understandable as \emph{stating} the truth of $\phi$ in $M$. We consider symbolic \emph{awareness} of $\phi$ as syntactically knowing the formula, and its semantics, i.e. \emph{being able} to, but not yet actually, evaluating it. An awareness \emph{knowledge base} $\Omega$ is a conjunction of any number of distinct symbolic awareness formulas.}

%Awareness  , i.e. to know if it holds. 
%As a paradox example, $\phi$ could actually happen to hold in $M$ while we are unaware of it, if we ignore the formula $\phi$ itself.

\smallskip
\noindent\textbf{Description Logics:} To implement awareness, we use the \emph{description logics} (DL) framework. %, since it has a computationally efficient implementation of classification tasks, i.e. deciding $M \vDash \phi\;$. Note that, if $M$ is a concrete instance of an object symbolic model, this just corresponds to deciding if it matches a known symbolic description $\phi$. 
DL are a family of logical formalisms to describe and reason about knowledge, and one of the mostly used to encode ontologies~\cite{Horrocks:2007ab}. In general, a DL is a subset of first-order logic (FOL)~\cite{baier2008principles}, trading some expressive power in order to have decidable and efficient decision procedures.

\smallskip
\noindent\textbf{Definition 4 (DL model):} {A DL \emph{model} is a set of declarative statements about three types of entities: \emph{concepts, roles and individuals. Concepts are named classes, individuals are their instances. A \emph{role} is a triple $<r, s, t>$, where s and t are the source and target entities, and can be of any kind. $r$ is a named binding, directed from source to target, i.e. $r(s)=t$  or, as a Boolean formula, $r(s,t)=true$.} In fact, roles correspond to binary predicates of FOL, and classes and individuals to unary predicates and constants, respectively.

\smallskip
\noindent\textbf{Generalized Reactivity:} For the automatic synthesis of a reactive controller, we leverage a polynomial time algorithm devised by Piterman et al.~\cite{Piterman2006} and based on principles of games theory. In this algorithm, the controller is obtained as the winning strategy of an infinite-duration, turn-based game of a specific type, called Streett games of depth one \cite{KESTEN200535}, or \emph{generalized reactivity} (GR(1)). In the modelled game, the system is confronted with the uncontrolled environment, as its adversary. A solution is found such that whatever next move the environment plays, the system will always have one available move preserving the satisfaction of its safety specifications. %To be applicable, the algorithm requires the ploblem to be expressed in  specifications to be compliant with a specific subclass of LTL called GR(1), representing \emph{assumptions} and \emph{conclusions}.

\noindent\textbf{Definition 5 (GR(1) games):} {\it A GR(1) game has to be expressed in the following conjunctive normal form: representing \emph{assumptions} on the environment and corresponding \emph{conclusions} about the system. \[\theta^e_0\wedge\bigwedge_{i=1}^{I_e}\square\phi^e_i\wedge\bigwedge_{k=1}^{K_e}\square\lozenge \phi^e_k\implies \theta^s_0\wedge\bigwedge_{i=1}^{I_s}\square\phi^s_i\wedge\bigwedge_{k=1}^{K_s}\square\lozenge\phi^s_k\]
where $\theta^e_0$, $\phi^e_{i}$ and  $\phi^e_{k}$ are Boolean formulas encoding assumptions on the environment possible start states, safety and liveness goals, respectively; similarly, $\theta^s_0$, $\phi^s_{i}$ and $\phi^s_{k}$ are the required system start states, safety and liveness goals, and $I_e$, $K_e$, $I_s$, $K_s$ are integer constants.}
%With a $strict$ implication\footnote{It makes no sense to solve the game by non satisfying both the assumptions and the guarantees}, the assumptions on the environment are required to guarantee system's safety ($\phi^s_{i}$) and liveness ($\phi^s_{k}$) goals, from its starting states ($\theta^s_0$). 

\begin{mdframed}
\textbf{Problem statement:} Given the model of the system as FTS, an LTL specification $\Phi$, and a knowledge base $\Omega$ encoding awareness about the unknown environment, design automatically a reactive controller to satisfy the specification using the knowledge base and the perception from the environment.
\end{mdframed}
%%%%%%%%%%%%%%%%%%%%%%%%%%%%%%%%%%%%%%%%%%%%%%%%%%%%%%%%%%%%%%%%%%%%%%%%%%%%%%%%%%%%%%%%%%%%%%
\section{Symbolic Awareness Modeling and Control}\label{sec:solution}
 Symbolic controllers for reactive motion planning normally rely on sets of symbolic abstractions modeling the available perception-module outputs ($U$) and the available control-module inputs ($Y$). Symbols in  $U$ and $Y$ are just labels, i.e. flat, unstructured models. % with context-free, self-explanatory semantics. 
 However, in general, these labels can abstract not only trivial entities, e.g. a color, but also non-trivial, structured entities, such as articulated environment objects or complex system behaviours, whose composing \emph{parts} or aspects are also relevant separately. 

\noindent\textbf{Knowledge Awareness Model:} We thus consider an extended symbolic model $\Lambda:= U^*\cup Y^*$, where $U^*\supset U$ and $Y^*\supset Y$, assuming to additionally include in it new symbols modeling shareable, partial properties of all non-trivial abstractions in $U$ and $Y$. %These are respectively \emph{partial environment inputs} and \emph{partial system behaviours} that are relevant in the operating context. 
We call the extended set $U^*$ the symbolic model of all {\it detectable} environment features and the extended set $Y^*$ the symbolic model of all implementable system \emph{behaviours}.   
We also extend the model with a union set of relations $\omega$ over the symbols in $\Lambda$, $\omega:=\omega^U\cup\omega^Y\cup\omega^{UY}$, where $\omega^U$ is a set of relations $U^*\mapsto U^*$ specifying the structure of the detectable features in terms of bindings to other features; $\omega^Y$ is a set of relations $Y^*\mapsto Y^*$ specifying the structure of the context behaviours in terms of other behaviours, and what's more important, $\omega^{UY}$ is the set of input-output \emph{relations} $U^*\mapsto Y^*$ specifying the environment's behavioural \emph{rules} of correct operation in the context environment. By construction, the  $\omega$ is also an ontology of the operating context, encoding \emph{knowledge awareness} of its known (since they are modeled) entities and semantic rules. We assume $\omega$ expressed as DL statements, and leverage its computationally efficient implementation of the \emph{classification} task, i.e. deciding if a concrete symbolic input $M$ matches a known symbolic description $\phi$. In fact, this in turn will correspond to the decision procedure for $M \vDash \phi\;$, where $M$ is an assignment to a subset of the input symbols in $U^*$, and $\phi$ a subset of the relations in $\omega$ modelling an environment object we are aware of. If $M \vDash \phi\;$ evaluates true we know that the detected \emph{entity} $M$ is an instance of object $\phi$.

\noindent\textbf{Awareness LTL specifications:} We take the set of \emph{roles} in Description Logic (DL), and translate this knowledge base into a set of LTL specifications in order to incorporate it into the controller symbolic computation. Transposition of a DL model $\omega$ into a set of LTL formulas is immediate, as long as the relations we included in $\omega^U$ and $\omega^Y$ are by definition unidirectional bindings between symbols at different levels of abstraction, i.e. \emph{sub-classing} relations between a shareable property symbol and a sharing property symbol. These are then equivalent, and can be directly mapped, to a set of logical implications between the involved symbolic variables: if symbol $a$ is a subclass of symbol $b$, the relation is logically encoded as $a \implies b$.   
Relations in $\omega^{UY}$, i.e. across $U^*$ and $Y^*$ symbols, encode a behavioural requirement of an abstract action $y \in Y^*$ to be implemented as \emph{next}, in response to the current detection of an abstract input in $x \in U^*$. This requirement is then also mapped into an equivalent LTL implication between the involved symbols: $x \implies \bigcirc y$. We call $\sigma$ the set of LTL formulas obtained from the translation of the semantic relations in $\omega$. Note that, this can be also done automatically by a compiler, in order to automate this step.

%\section{Symbolic Controller Synthesis}
\noindent\textbf{Symbolic Controller Synthesis:}\label{sec:synthesis} Synthesis of a \emph{context aware} controller is obtained using the Tulip tool~\cite{wongpiromsarn2011tulip} for reactive controller synthesis from GR(1) specifications. It requires a FTS as model of the system spatial motion abilities, and a set of LTL specifications.%Specifically, the system LTL safety section of the specifications also the set of LTL implications encoding the $\sigma$ knowledge base. 
We incorporate into the system the \emph{awareness} of the operating environment, enabling it to recognize and {\it understand} known objects, and to deal with their partial symbolic features, thus implementing gradual reactions. Specifically, assuming no other system or environment safety requirement is given, in the GR(1) game formulation we will have $\bigwedge_{i\in I_s}\phi^s_i=\sigma$ while $\bigwedge_{i\in I_e}\phi^e_i$ will be empty. The liveness goals $\phi^{s/e}_i$ can be any Boolean formulas built upon symbols in $\Lambda$. Since these are always specific to the considered application, e.g. what destination to reach or other particular task to accomplish infinitely often, they have to be specified separately, on a case by case base.

%%%%%%%%%%%%%%%%%%%%%%%%%%%%%%%%%%%%%%%%%%%%%%%%%%%%%%%%%%%%%%%%%%%%%%%%%%%%%%%%%%%%%%%%%%%%%%
\section{CASE STUDY}
\label{sec:casestudy}
We apply our approach to a case study of an autonomous car driving in a road with traffic signs but no obstacles, depicted in Fig.~\ref{fig:scenario}.
%The road signalization system appears to be a complex system that is not fully coherent since the surface properties of the signalling devices often match partially with the corresponding (user-required) action properties, which makes it a perfect context to exploit awareness-enhanced symbolic computation.
%
%\noindent\textbf{Car model:} A simple ego-car model of an autonomous vehicle has been designed for the purpose of this case study as a

The car is modelled with a
non-deterministic FTS depicted in Fig.~\ref{fig:fts}, which represents the car moving along five cells $location\; i$, $i\in\{0,1,2,3,4\}$. In each cell, the car can either be moving or stopping, where a stop can be implemented in one single cell (hard stop action) or by slowing down in order to smoothly stop \emph{n} cells ahead.
%Simple straight motion up to the target is assumed since complex route planning is not the focus of this study. 
%The space model has then been designed as a closed loop so that
We assume the car progresses in an infinitely repeated sequence of five cells paths, with a stop sign at cell $location\; 0$. The car liveness goal is to reach infinitely often next stop sign and move on again.

\begin{figure}[t]
    \centering
    \includegraphics[width = \columnwidth]{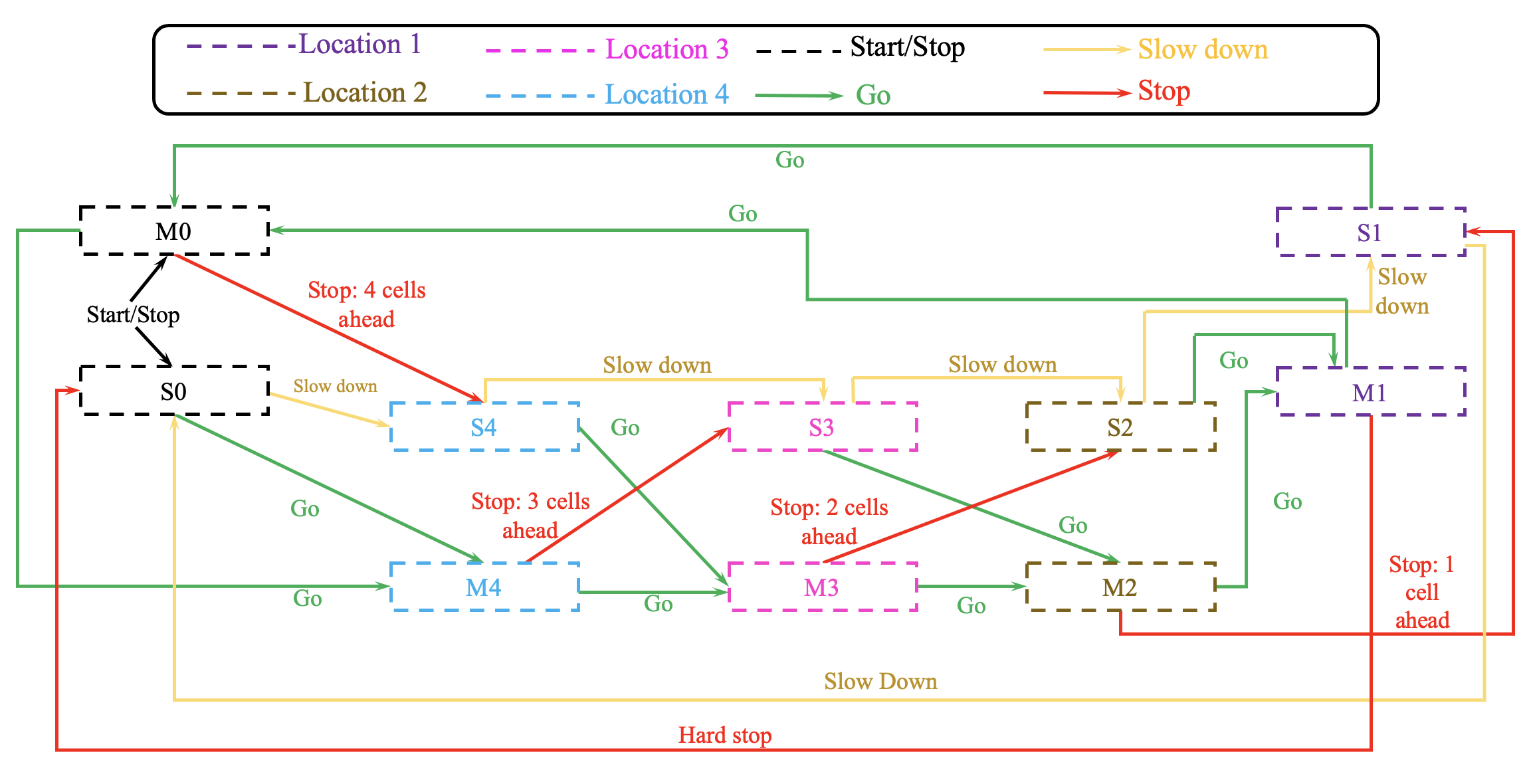}
    \caption{FTS model of the car. Dashed boxes correspond to the states, modelling both cells $location\; i$, $i\in\{0,1,2,3,4\}$ and the actual moving or stopping state.
    %A moving state is an abstraction of a car that is accelerating up to max speed, while 
    The stopping state represents the car decelerating (S4, S3, S2, S1) or hard stopping (in S0 only). A stop sign is simulated to be present only at $location\; 0$.
    %The car progresses in an infinite loop starting/ending at S0, if it managed to stop at the sign, or M0, if not. 
    The FTS is non-deterministic, always allowing the car to go from a moving state to stopping and vice versa.}
    \label{fig:fts}
\end{figure}

\noindent\textbf{Knowledge awareness model:} We have selected a representative subset of EU traffic regulation signs and analysed them to extract a semantic knowledge base as a standard \emph{OWL2} ontology~\cite{semanticweb-owl}.
Fig.~\ref{fig:onto} shows the
entities of the
\emph{traffic signs} ontology defined in the Protegè tool \cite{Musen15}. It includes concepts modelling the signs as environment objects, the features as abstractions of perception inputs, and actions as behavioural prescriptions of the signs.
\begin{figure}[t]
    \centering
    \includegraphics[width=0.9\columnwidth]{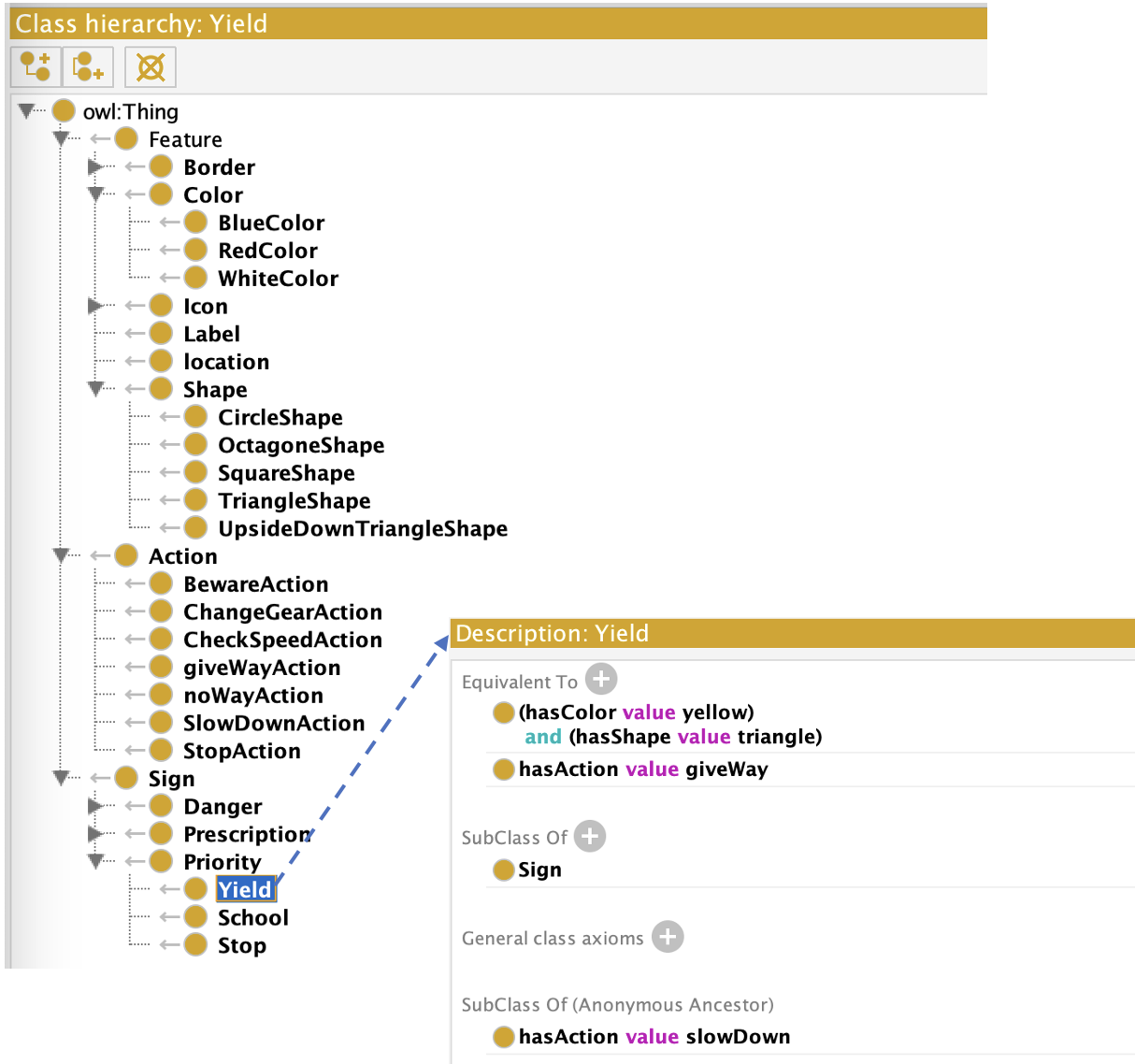}
    \caption{Entities of the \emph{traffic signs} ontology from the Protegè tool \cite{Musen15} with an example of simple semantics for a ``yield'' sign.}
    \label{fig:onto}
\end{figure}
%A classic symbolic inputs model would have flat symbolic abstractions for the road objects and their features. For instance, a \emph{colour or shape} detected by the sensor, a \emph{location} in the discrete space, an \emph{obstacle} or a \emph{traffic sign} (if modelled as trivial features). The symbolic outputs model commands are related to motion abilities, like \emph{go, slow down, stop}, etc. 

This ontology has modelled \emph{concepts} including the sign features, e.g.  \emph{colours or shapes}, that are shareable (partial) input features; the \emph{``Signs''}, which are non-trivial objects, together with some group \emph{classes} of sign types, modelling their categories. The set $U^*$ of the detectable features for our case study holds symbolic abstractions for each of these concepts.  The ontology has also modeled the \emph{Actions} prescribed by the signs: these will be modelled as symbolic abstractions in the set $Y^*$ of the system-required behaviours. %The syntax (structure) and semantics (meaning) of each traffic sign is sparsely encoded in the bindings between these model entities. 
Semantic bindings between Signs, Features and Actions are defined in the ontology by named \emph{relations}: \emph{hasColor} (e.g. to bind the ``red'' feature to the ``stop'' sign), \emph{hasShape}  (e.g. to bind an ``octagonal'' shape to a ``stop'' sign), and \emph{hasAction} (e.g. to bind ``halt'' and ``giveWay'' actions as prescriptions for the ``stop'' sign). 
\begin{figure}[t]
    \centering
    \includegraphics[width=\columnwidth]{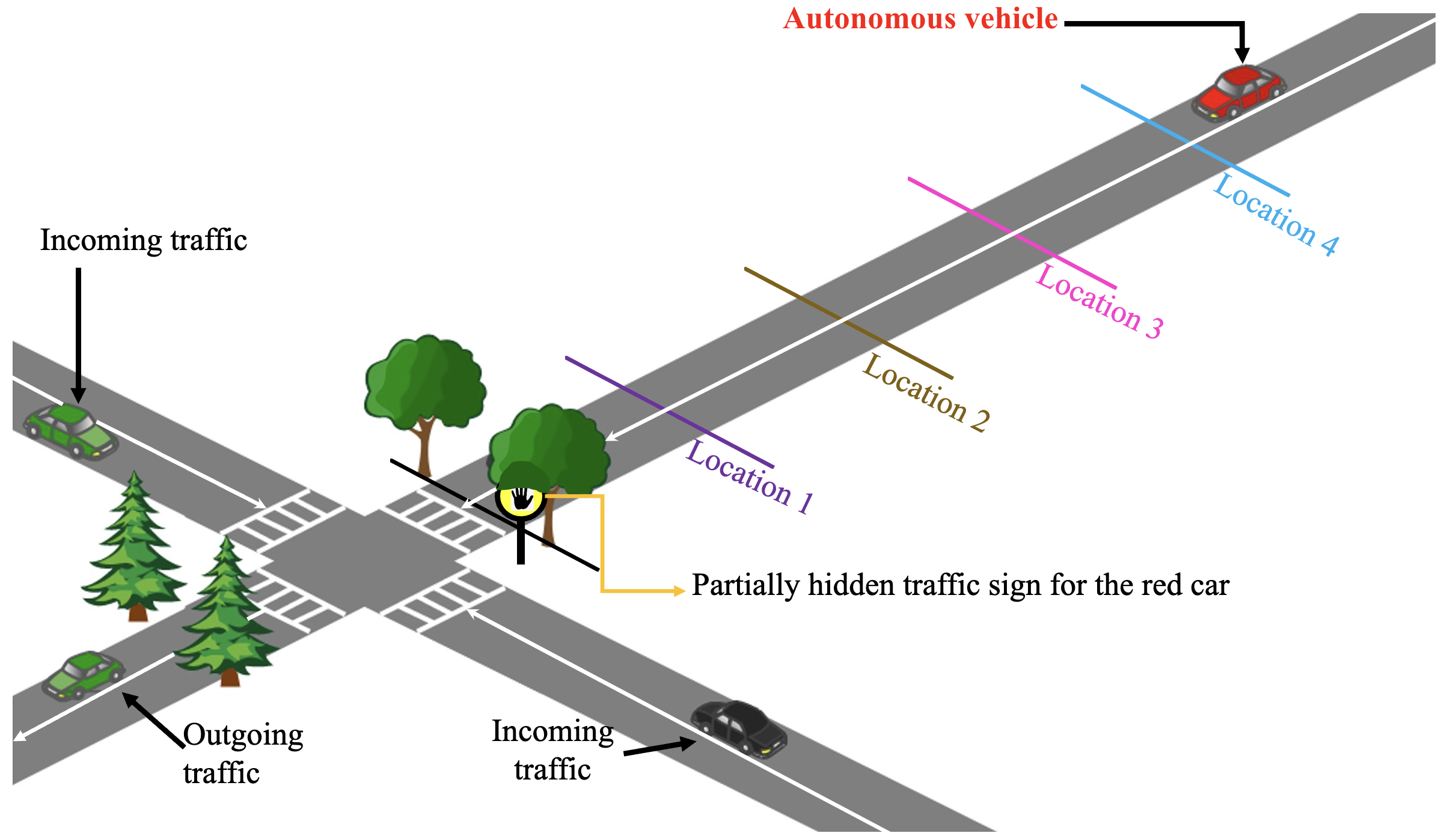}
    \caption{The tested scenario}
    \label{fig:scenario}
\end{figure}

\noindent\textbf{LTL specifications extraction:} To understand LTL extraction from DL semantic relations, consider as an example the case of the \emph{yield} traffic signal, which in the traffic signs ontology is modelled as an object of class \emph{sign} with \emph{yellow}, and \emph{triangular} features, whose prescribed action is to \emph{give way} to other cars. Assume also that the generic ``sign'' class has a semantic binding to a \emph{slow\_Down} action, to pay attention to whatever is the upcoming signal. This is encoded as the following DL role assertions:\begin{quote}
   \it (hasColor yield yellow) $\wedge$ (hasShape yield triangle) $\wedge$ (hasAction yield giveWay) $\wedge$ (isSubClass yield sign) $\wedge$ (hasAction sign slowDown)
\end{quote} %where yield, yellow, triangle and giveWay are symbolic vars and hasColor/hasShape/hasAction are binary relations. 
The yield sign LTL specification (compacted) then results as:
\begin{center}
    $ sign \wedge red \wedge triangle \implies yield $\\
    $ yield \implies \bigcirc\; giveWay $\\
    $ sign \implies \bigcirc\; slowDown $\\ 
\end{center}
%\Sadegh{Is it yield sign or yield sign? Are we using the right word here?}
In fact, since ``yield'' is a subclass of ``sign'', and subclass relations become an implication, the semantic relation associating slow\_Down action to sign will be inherited by the ``yield'' system specification.

\noindent\textbf{Early Reactions:} As shown in the above example, the generic ``sign'' input is a partial/incomplete input feature, denoting detection of a sign of still unknown type. However, it already triggers its own \emph{early} re-action in the controller, i.e. to immediately start to progressively ``slow\_down''. Conversely, the ``giveWay'' action, bound to the actual ``yield'' sign detection, will be implemented only later, at full sign detection. 
%The \emph{knowledge base} extracted from the traffic signs ontology semantic relations and put into the controller synthesis as LTL specifications, results in a controller with \emph{awareness} of the entities that are \emph{possible} in the interaction context, i.e. with a \emph{semantic understanding} of traffic signs structure and meaning, including also the ability to react to the sign partial features.
This ability to anticipate part of required actions results in \emph{safer and smoother} system behaviours with respect to a classic symbolic controller, which is our primary objective. The outcomes of the simulation run presented in the following show that this has been systematically achieved. 
\begin{figure}[t]
    \centering
\includegraphics[width=\columnwidth]{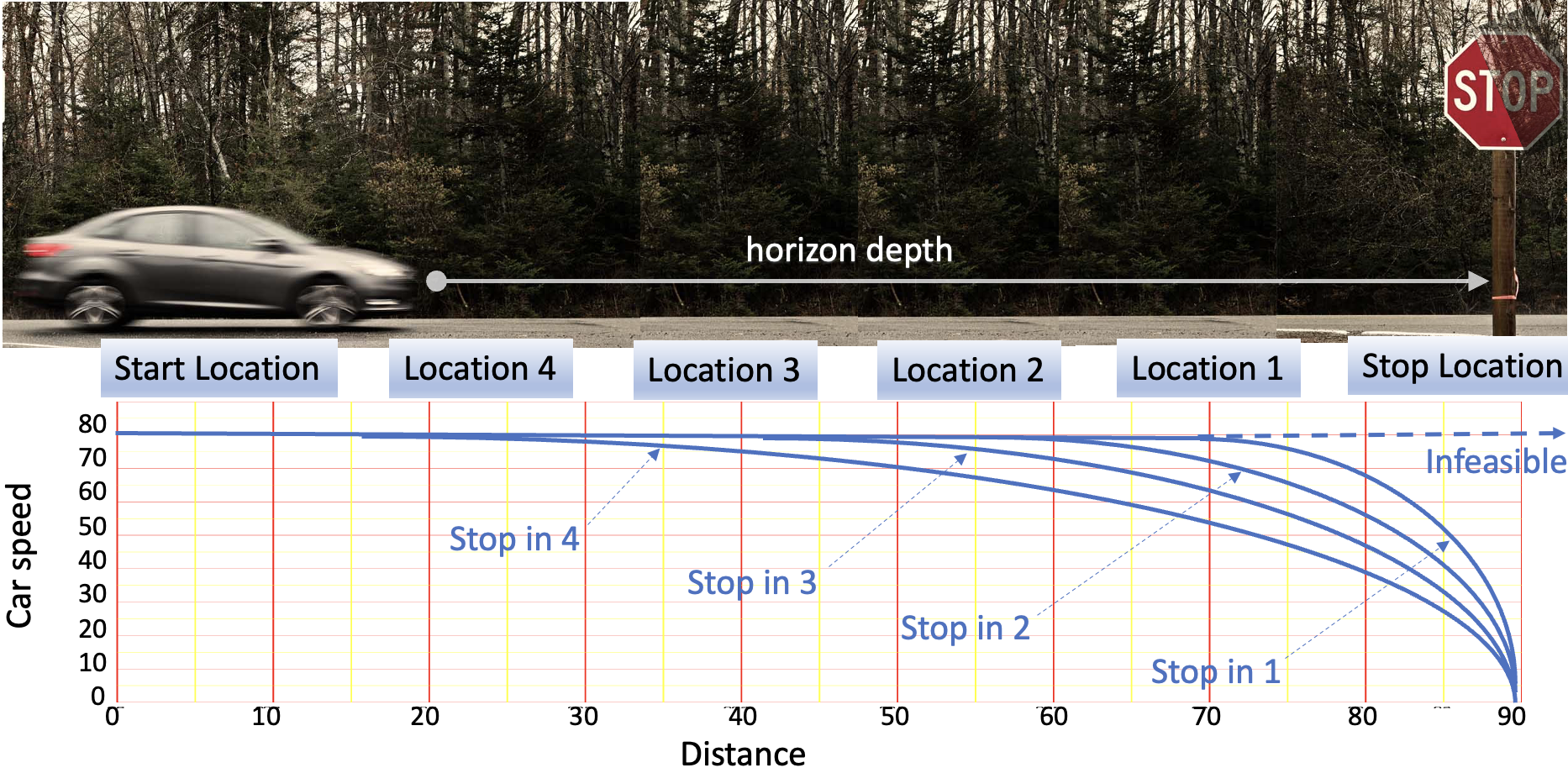}
    \caption{Smoothness of the car trajectory with respect to early semantic understanding of collected partial input features, resulting in early detection and planning of a stopping behaviour. A ``Stop in 4'' trajectory means the car detects (understand) it has a stop sign four cells ahead of it, and is then able to implement a smoother stop as four progressive slow-down control actions. }
    \label{fig:smooth}
\end{figure}

\noindent\textbf{Simulation setup:} We tested the car behaviour in an unknown environment, assuming a partially visible stop traffic sign appears at the horizon line, as depicted in Fig.~\ref{fig:scenario}, as a representative scenario whose results can be easily extended to other traffic signs. We set up the simulation to run a large number (10,000)  of statistically varied, five cells sequences of random environment inputs. %This to  incremental perception of partial features of an actual sign, present at the end of each path. The large number (ten thousand) of random paths per simulation run, with randomly varied, partial inputs sequences simulating environment uncertainty,
This allowed us to confidently achieve a statistical figure of the ability of the knowledge-aware controller to react to input uncertainty and to contrast it to existing approaches. %Symbolic environment random input traces of length five for the paths have been generated with three different randomization strategies, to produce a wide statistical exploration of the legal environment behaviors for that scenario. The simulation aims at contrasting existing approaches to our approach performance in terms of its ability to understand the environment and implement the required behaviour for a larger set of the varying, unknown sequence of symbolic environment inputs.  

\noindent\textbf{Multiple Controller strategies:} Three instances of traffic signs controllers have been synthesized, in order to compare their performances: base, perception-tree and knowledge-awareness based.  The \textbf{base controller} is used for reference and is a plain symbolic controller, not leveraging perception refinement nor knowledge awareness. It can only detect and react to traffic signs when they are visible clearly in full. The second implements a symbolic \textbf{perception-tree controller}, that is, it assumes that the perception layer will always produce increasing levels of symbolic object details precisely in a well-known sequence. This controller can then react early to partial features detected along the perception tree. The third is an \textbf{awareness controller}, incorporating a fragment of the knowledge base for traffic signs, semantically describing the environment inputs without constraints on their perception sequence. As a consequence, this controller reacts to partial/incomplete set of features for a larger fraction of the explored environmental input sequences. All controllers have been automatically synthesized with Tulip, using the \emph{Omega} resolver for GR(1) games. To enable a fair comparison with the perception-tree controller, all controllers assume persistent and incremental environment inputs. %\andrea{This assumption has been encoded into the environment LTL safety specification $\phi^e$, defining \emph{legal} input sequences.} 
%The environment has no liveness goals, since its inputs are externally generated in the simulation. 

\noindent\textbf{Random trace profiles:} Three different random \emph{profiles} have been used to simulate uncertainty of the environment input sequences. \textbf{Profile \#1} models features appearing with a linear probability distribution function (pdf) over their distance from the car, starting from 50\% visibility chance, at max horizon distance, to full feature visibility (100\%), at the sign location. \textbf{Profile \#2} randomly chooses the location where the first partial feature detection happens, and then adds a new random feature detail at each successive location. \textbf{Profile \#3} also allows all partial features to be randomly spaced from each other. %All the input simulation profiles have been applied to the three synthesized controllers (base, tree or awareness).
\section{EVALUATION}\label{sec:evaluation}

We measured performance primarily in terms of safety, and then the smoothness of the car's implemented behaviours.  Safety is of utmost importance and is measured by counting the frequency of behaviours where the car fails detect and stop at the ``stop'' sign. These events are statistically possible due to the uncertainty of the environment information modelled in the simulated traces, and correspond to cases where the detection of the sign happens too late, i.e. at the sign location or later. They are shown as \emph{infeasible} behaviours and should be as low as possible. Secondly, while being safe, it is highly desirable to detect the stop sign also as early as possible, in order to have more space to implement a smoother stopping behaviour, i.e. starting with $n$ cells of anticipation (see Fig.~\ref{fig:smooth}). This has been counted in columns \emph{stop in $n$}. Since stopping within four cells space is smoother than stopping in three, and so on, the smoothness performance evaluation shall privilege higher values in the rightmost columns.

Fig.~\ref{fig:pdf} shows the performance comparison of the three control strategies with input traces profile \#1: both perception-tree and awareness-based controllers significantly improve safety over the base controller, and the awareness controller shows also superior smoothness performance with respect to the perception-tree controller, having 10\% higher number of stop manoeuvres started with three cells of anticipation.
%Note that having a cap on the total number of tests, column heights must balance with each other, so a higher count in column stop at $n$ inevitably results in lower counts for the previous ones, and vice versa.

\begin{figure}[ht]
    \centering
\includegraphics[width=\columnwidth]{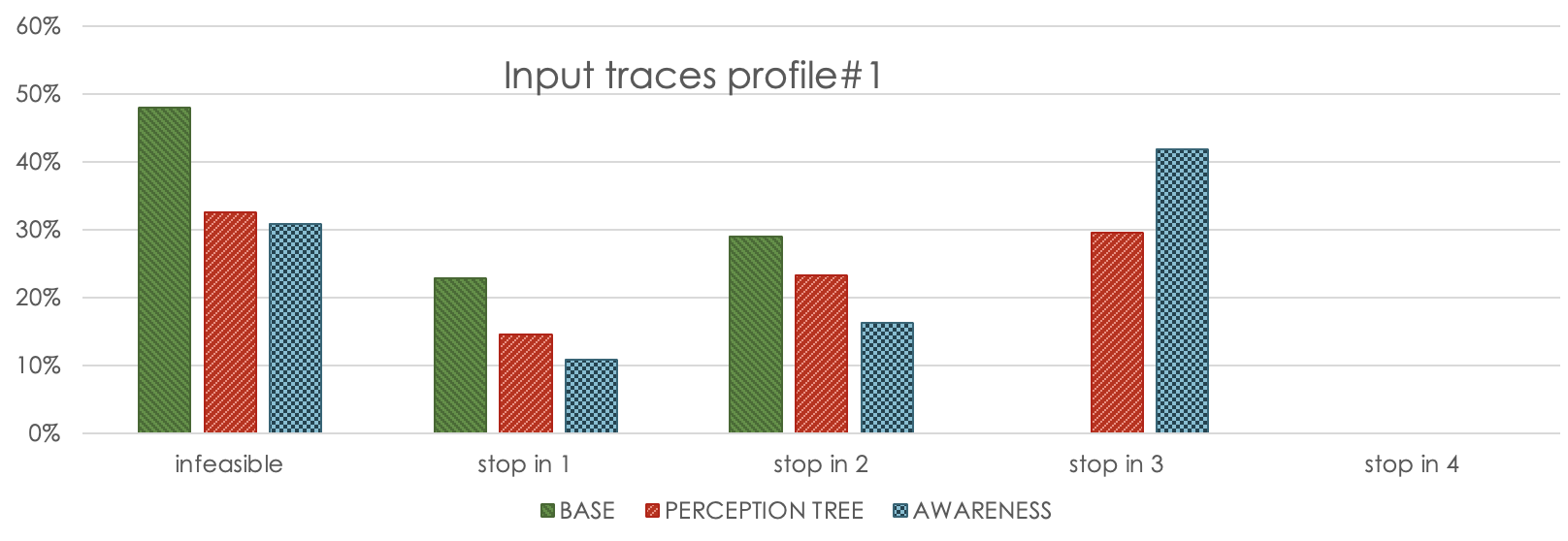}
    \caption{Random sequence of increasingly visible partial features. in these charts, X-axis column correspond to early stop detection of type ``stop in n'' or infeasible. Y-axis is the counted fraction of observed behaviours for each x-axis column type.}
    \label{fig:pdf}
    \centering
    \includegraphics[width=\columnwidth]{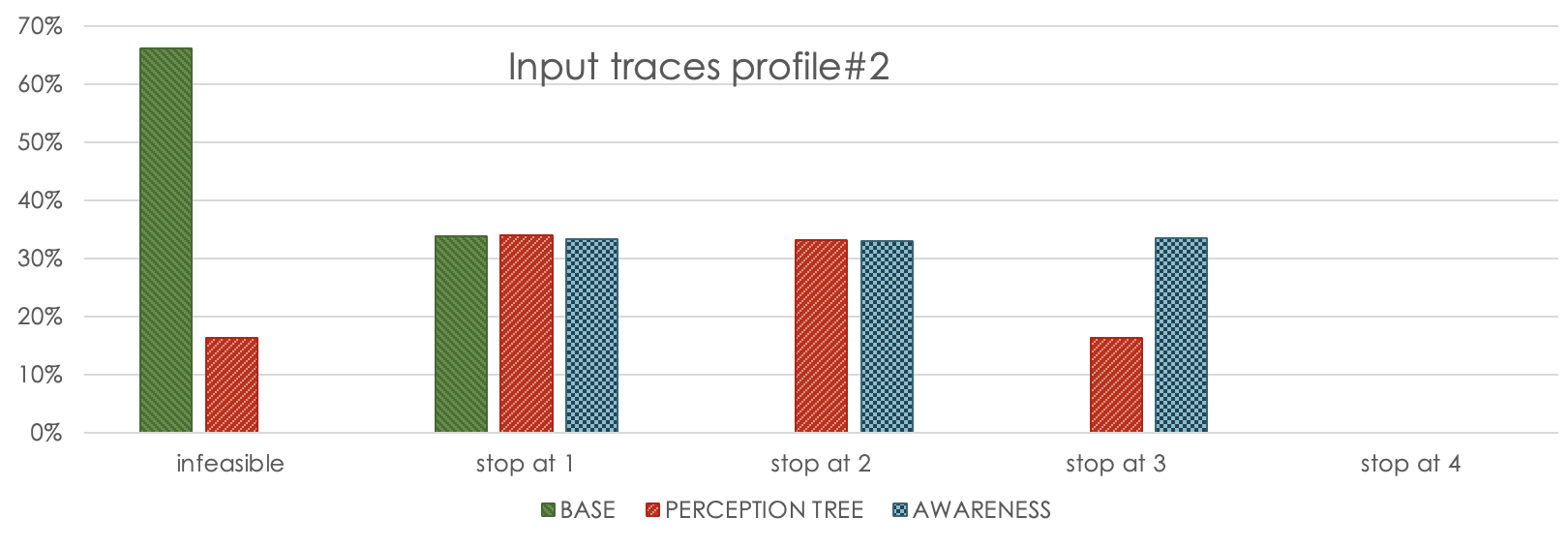}
    \caption{Random sequence of closely consecutive partial features.}
    \label{fig:strict}
     \centering
    \includegraphics[width=\columnwidth]{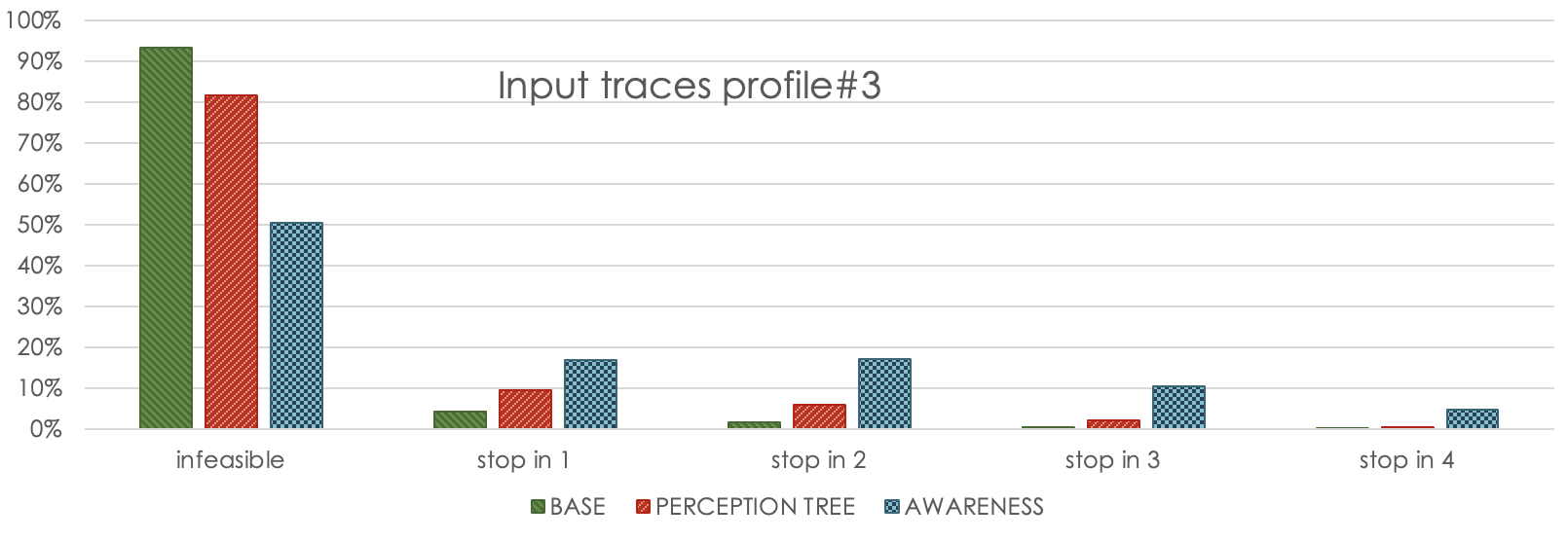}
    \caption{Random sequence of loosely consecutive partial features.}
    \label{fig:loose}
\end{figure}
Fig.~\ref{fig:strict} shows the controller performance comparison for input traces profile \#2: The base controller is able only to detect the sign in the very last useful space location in 35\% of the tests while missing it for the remaining 65\%. Perception-tree improves significantly over this, but the awareness controller performs dramatically better than both, having not missed a single stop behaviour.
%doubling the smoothness performance with respect to the perception-tree controller in terms of early stops with 3 cells of anticipation, and having a comparable performance for stops in 2 and 1 cells distance. 

Fig. \ref{fig:loose} shows the performance comparison with the input traces profile \#3: The base controller has always had the worst performance, having about 90\% of traces ending up as missed stops, while the perception tree reduces it to 80\% and the awareness-based controller again improves significantly over the perception tree, reducing it to 50\%. It is also neatly superior in smoothness of the stop behaviours.
%and the only able to implement a relevant number of early stopping behaviours starting even four-cells ahead.

\section{CONCLUSIONS}
\label{sec:conclusion}
%elaborate on the importance of the work or suggest applications and extensions. 
We presented a technique to incorporate semantic knowledge awareness of the operating context into the synthesis of symbolic controllers for reactive systems autonomous driving in unknown and uncertain environments.
%as no constraints are required on the input dynamics. %as refinement is now implicitly and inherently driven by any dynamics of the symbolic inputs. (can we restructure the sentence!!)} %This also allows to reveal and possibly recover from misleading input updates, which was not possible with perception-trees. 
%We incorporate symbolic knowledge awareness of environmental features into the automated synthesis of reactive controllers, using also state-of-the-art scaling algorithm. 
We applied the presented technique to a case study simulating a car that reacts to random partial inputs in an unknown road environment. The awareness-enhanced controller showed significant performance improvements in terms of both safety and smoothness of behaviours  compared to existing alternative control techniques also supporting perception refinement, proving that knowledge awareness modeling subsumes those approaches and is a valid and promising direction to further develop the design of safe reactive controllers for autonomous systems.
%contrasted the simulations results of the  with those of basic and perception-tree-based controllers, 
%. %\andrea{, this automated technique could be coupled with machine learning (ML) specialized algorithms, applied to capture the knowledge awareness models of any large, real world domain of interest, in order to have automated real time controller synthesis available for any context of interest.}
%With large enough data records available, state-of-the-art machine learning (ML) algorithms can be applied to capture any real-world domain ontology that is not yet available and extract it into an awareness knowledge base. 
With the presented technique, it is also possible to re-use the knowledge awareness of a context into the automated synthesis of safe-by-design reactive controllers for multiple specific applications. As a new line of research we consider for future development, it is also possible to enable a knowledge aware CPS to learn and update its knowledge base itself, through the continuous analysis of its own \emph{experience}.  
%In contrast to perception refinement approaches, knowledge awareness supports incremental updates of object detection without requiring \emph{additive} symbolic perception or \emph{persistence}, allowing the controller to understand a much larger set of unpredictable environmental behaviours, including incomplete object descriptions due to missing features. 
%and also to reveal and possibly recover misleading input updates
 %each refined feature is a symbolic model of a whole predefined \emph{ordered list} of \emph{additive} feature details of a corresponding object. As a consequence, different feature orderings or incomplete sets of additive features would be simply not understood by the controller.

%%%%%%%%%%%%%%%%%%%%%%%%%%%%%%%%%%%%%%%%%%%%%%%%%%%%%%%%%%%%%%%%%%%%%%%%%%%%%%%%
%\section*{ACKNOWLEDGMENT}
%Dear editor, this work is the result of our original research, and contributes to the dissemination of the Horizon Europe EIC project SymAware. It is the first time that cognitive logic is integrated into the actual synthesis of a reactive controller. All developed research materials are available on request to guarantee that presented results are verifiable and reproducible. We would be glad if you consider it for publication in IEEE Control Systems Letters. 
%%%%%%%%%%%%%%%%%%%%%%%%%%%%%%%%%%%%%%%%%%%%%%%%%%%%%%%%%%%%%%%%%%%%%%%%%%%%%%%%
\bibliographystyle{IEEEtran}
\bibliography{lcsscdc23}
\end{document}